\documentclass{article}
\usepackage{spconf}

\usepackage[title]{appendix}

\usepackage[latin1]{inputenc}
\usepackage{amsmath,amsthm}
\usepackage{amsfonts}
\usepackage{amssymb}
\usepackage{graphicx}
\usepackage{mathcomSTEv4}
\usepackage{color}

\usepackage{tikz}
\usepackage[outline]{contour}

\tikzstyle{int}=[draw, fill=blue!10, minimum height = 1cm, minimum width=1.5cm,thick ]
\tikzstyle{sint}=[draw, fill=blue!10, minimum height = 0.5cm, minimum width=0.8cm,thick ]
\tikzstyle{sum}=[circle, fill=blue!10, draw=black,line width=1pt,minimum size = 0.5cm, thick ]
\tikzstyle{ssum}=[circle, fill=blue!10,draw=black,line width=1pt,minimum size = 0.1cm,inner sep=0pt]
\tikzstyle{int1}=[draw, fill=blue!10, minimum height = 0.5cm, minimum width=1cm,thick ]
\tikzstyle{enc}=[draw, fill=blue!10, minimum height = 2.7cm, minimum width=1cm,thick ]
\tikzstyle{int}=[draw, fill=blue!10, minimum height = 1cm, minimum width=1.5cm,thick ]

\usetikzlibrary{patterns}

\newtheorem{remark}{Remark}

\begin{document}

\title{
The statistical dictionary-based \\
string matching problem
}
\name{M. Suri and S. Rini  
}
\address{Electrical and Computer Engineering Department, \\
	 National Chiao Tung University (NCTU), 	Taiwan	
}

%
%

\maketitle

%
\begin{abstract}
%
In the Dictionary-based String Matching (DSM) problem, a retrieval system has access to a source sequence and stores the position of a certain number of strings in a posting table.
When a user inquires the position of a string, the retrieval system, instead of searching in the source sequence directly, relies on the the posting table to answer the query more efficiently. 
%
%
%
In this paper, the  Statistical DSM problem is a proposed as a statistical and information-theoretic formulation of the classic DSM problem in which both the source and the query have a statistical description while the strings stored in the posting sequence are described as a code. 
%
%
Through this formulation, we are able to define the efficiency of the retrieval system as the average cost in answering a users' query in the limit of sufficiently long source sequence.
%
This formulation  is used to study the retrieval performance for the case in which (i) all the strings of a given length, referred to as $k$-grams 
, and (ii) prefix-free codes.
%
%
\end{abstract}
%
%
%

\section{Introduction}
\label{sec:intro}

Let us define a Dictionary-based String Matching (DSM) problem is defined as follows.
The \emph{retrieval system} has access to a \emph{source sequence} and constructs  a \emph{posting table} in which it stores the position of a set of source substrings, referred to as a \emph{posting code}.
More precisely, each row in the posting table contains the \emph{posting list} corresponding to a given codeword, consisting  of a list of positions in which the codeword appears in the source sequence, as selected by the \emph{posting function}.

At a later time, a user submits a string to the retrieval system, termed a \emph{query}, and the retrieval system is tasked with providing the positions of the query in the source sequence, called \emph{matches}.  
If the query does not appear in the source sequence, a empty message is returned to the user. 
If the retrieval system is not able to retrieve some of the matches, an error message is returned.

We consider a variation of the classic DSM in which we assume that (i) the source sequence and the query have a statistical description and that (ii) the cost of a query is proportional to the product of the length of the entries in the posting table visited by the retrieval system in answering a query. 
Under these two assumptions, we consider the problem of designing the posting code that  minimizes the expected cost of retrieving the positions of a query, from the information in the posting sequence in the limit of an infinitely long source sequence.
We term this problem as the Statistical DSM (SDSM). 
We are interested in the study of the SDSM as we wish to determine the ultimate information searching efficiency in the posting table. 
As such, this paper represents a stepping stone toward  the development of a universal and dynamic SDSM in which the source sequence is any stationary sequence while the query distribution is unknown at the retrieval system.
%
%
%
%

\noindent
{\bf Literature Review:}
%
The DSM problem has been studied in a number of context and modeled through various assumptions so that a vast literature is available on the topic. 
To the best of our knowledge, no formulation has explicitly considered either the distribution of source and queries, or the performance in the limit of large source length. 
In the information retrieval context,  the DSM problem is referred to as ``inverted index'' problem and the concern is with the 
respect to the memory required to store the  entries of the posting table \cite{baeza1999modern}.
The distribution of the queries is used in \cite{baeza2003three} to design a  three level memory organization for a
search engine inverted file index.
In computation lingustics and natural language processing, the DSM problem has been studied to determine robust retrieval methods \cite{baeza1998fast,mihov2004fast} to search in a text affected by errors. 
In the context of genomics, bioinformatics and computational biology, the problem is sometimes referred to as off-line or indexed pattern matching: here the focus is on the retrieval of sequences  that approximatively match the given query \cite{apostolico1997pattern,navarro2001indexing}. 
More generally, the source distribution implicitly appears in the literature concerned with the compression of the entries of the posting table, such as \cite{anh2005inverted}, \cite{yan2009inverted}  and   \cite{chen2010inverted}, 
%
%
%

\noindent
{\bf Contributions \& Organization:}
The remainder of the paper is organized as follows


\noindent
$\bullet$ {\bf Sec. \ref{sec:Problem Formulation}- Problem formulation:}
We propose formulation of the SDSM problem which accounts for source and query along with the cost of accessing the entries in the posting table. 
We define the efficiency of a retrieval system as the expected cost of retrieving a query in the limit for large source length.
Through this performance measure, we formulate an optimization problem that helps us determine the code with the optimal memory utilization. 
%

%

\noindent
$\bullet$
{\bf Sec. \ref{sec:examples}- Relevant examples:}
To validate the propose model, we study in detail the performance of two retrieval system: one storing (i) $k$-grams, all possible  source sequences of length $k$  and (ii) prefix-free codes, codes in which no codeword is a prefix of another codeword.
For illustrative purposes, we consider the simple case of binary i.i.d. source and query distributions. 
%
%
	
\noindent
$\bullet$
{\bf Sec. \ref{sec:Numerical Evaluations}- Numerical Evaluations:} we numerically investigate the design of the optimal code for the case of binary i.i.d. source and query distributions. 
	
%

%
\medskip

\noindent
\emph{Notation:}
With $\xv= [x_1, \ldots, x_N] \subseteq \Xcal^N$ we indicate a sequence of elements from $\Xcal$ with length $N$.
The notation $\xv_i^j$ indicates the substring $[x_i, \ldots, x_j]$.
Given the sequence $\xv$, $l(\xv)$ indicates the length of the sequence $\xv$,  $w(\xv)$ indicates the Hamming weight, respectively.
%
%
The notation $\av.\bv$ indicates the vector concatenation operation.
%
The notation $\av \preceq \bv$ indicates that  $\av$ is a substring of $\bv$.
%
%
Let $\Pcal(\Xcal)$ indicate the power set of $\Xcal$.
Define $\xo=1-x$.
%
%
%
%
%

\medskip


\section{Problem Formulation}
\label{sec:Problem Formulation}
%
The  SDSM problem is comprised of a \emph{source sequence}, a  \emph{retrieval system}  and a  \emph{user}. 
The source sequence is defined as the random sequence $X^N$ with support $\Xcal^N$  and distribution $P_{X^N}(\xv)$ and let the \emph{query} be defined as the Random Variable (RV) $Q$ with support  $\Pcal(\Xcal) \setminus \emptyset$ and with distribution $P_{Q}(\qv)$.
%
%
%
A retrieval system is comprised of a \emph{posting code}, a \emph{posting table}, a \emph{storing function} and a \emph{retrieval function}.
A posting code of size $M$ is defined as the set $\Ccal=\{\cv_i\}_{i=1}^M$ with $\cv_i \in \Pcal(\Xcal^N)$.
The set of \emph{source matches}  of the codeword $\cv_i$ is the set $\Ical(\cv_i) = \{W_m(i)\}_{m \in \Nbb}$ for which
\ea{
X_{W_m(i)}^{W_m(i)+l(\cv_i)}=\cv_i, \ \ \ \forall \ i \in [1:M].
\label{eq:posting condition}
}
%
The \emph{posting list} of the codeword $\cv_i$ is defined as $\Tcal(\cv_i,X^N)$ and is such that $\Tcal(\cv_i,X^N) \subseteq \Ical(\cv_i)$.
The \emph{posting table} is defined as the tuple $T(\Ccal,X^N)=\{\Tcal(\cv_i,X^N) \}_{i=1}^M$.
%
%
%
%
%
The  \emph{storing function} is the mapping  $f_S(\Ical(\cv_i),X^{N})$ which produces the posting list from the set of source matches for each codeword $\cv_i$, i.e.
\ea{
f_S: \ \ \Ical(\cv_i) \goes \Tcal(\cv_i,X^N) \subseteq \Ical_{i}.
}
%
%
%
%
%

A user provides a \emph{query} $Q$  with distribution $P_Q(\qv)$ to the retrieval system: upon receiving a query $Q$,  the retrieval system  produces a \emph{covering} of length $V$ of the query $Q=\qv$, defined as the tuple $\Scal(\qv)=\{\cv^{(V)},\nv^{(V)}\}$ 
such that 
\ea{
\qv_{n_{(i)}}^{l(\cv_{(i)})}=\cv_{(i)}, \ \ \ \forall \ i \in [1:V].
}
%
%
%
%
%
%
If a covering of the query does not exists, a \emph{retrieval error} is declared. 
Once a covering is produced, the retrieval system fetches the position of the codewords in the covering from the posting table.
%
Finally, the \emph{retrieval function}, $f_R$, is the mapping
\ea{
f_R: \ \ \{ \Tcal(\cv_{(i)},X^N), \  \cv_{(i)} \in \Scal(\qv) \}_{i=1}^V \goes \mv \subset \Rbb^N,
}
where $\mv$ is such that $X_{m_i}^{l(\qv)}=\qv$
for all $m_i \in \mv$.

The average size of the posting list and the average size of the posting table are defined as
{
\small
\ean{
	\Ebb\lsb |\Tcal(\cv_i,X^N)| \rsb &= \Ebb[l (\Tcal(\cv_i,X^N))] \\
	\Ebb\lsb |\Tcal(\Ccal,X^N)| \rsb &= \sum_{i=1}^M \Ebb\lsb |\Tcal(\cv_i,X^N)|\rsb,
}
}
respectively.
If a covering for the query exists, than the cost of a covering $\Scal(\qv)$ is defined as
\ean{
c\lb \Scal(\qv) \rb & =\log \prod_{i=1}^V \labs \Tcal( \cv_{(i)}, X^N)  \rabs
=\sum_{i=1}^V \log |\Tcal( \cv_{(i)},X^N)|.
}
If a covering for the query does not exist, than the cost of the query is infinite.
%
The \emph{minimum  expected cost} for a given  query $Q=\qv$,  $c(\qv)$, is defined as 
\ea{
\Ebb[c^*(\qv)]= \min_{\Scal(\qv)} \Ebb\lsb  c\lb \Scal(\qv) \rb \rsb.
\label{eq:min cost}
}

%

Finally, we are now  ready to state the optimization problem of our interest.
For given source, query distributions, and the size of the posting code,  the maximal efficiency $\eta$ in the SDSM problem is defined as 
\ea{
\eta=\min_{ \Ccal, \ f_S(\cdot)} \ \lim_{N \goes \infty}  \f {\Ebb[c^*(\qv)]} {\log N}.
\label{eq:definition efficiency}
}

\begin{remark}
The cost function in \eqref{eq:min cost} is chosen so as to approximate the complexity of finding the positions in the entries of the posting lists in $\Scal(\qv)$ corresponding to contiguous codewords in the covering. 
See \cite{buttcher2016information,baeza2004fast,baeza2005experimental}.
	
\end{remark}

\subsection{The Pre-fix free coded, Complete and Parsed (PCP) SDSM problem}
In the above formulation, the SDSM problem is presented in the greatest possible generality. 
In the following, we focus on a specific formulation of the SDSM problem, the Pre-fix free coded, Complete and Parsed (PCP) SDSM problem, which can be more readily analyzed.
In particular, we consider the case in which (i) the posting code is a complete pre-fix free code (see \cite{yeung2012first}), (ii) the posting table stores all the matches, (iii) queries are covered by non-overlapping codewords. 
%
%
%
%
While property (i) and (ii) are straightforward, for (iii), we resort to the following definition. 
A covering is defined as a \emph{parsing} if there  exists an $K$ such that 
$n_j+l(\cv_{(j)})=n_{j+1}$  
for $j\in [1:K-1]$,  while 
\ea{
n_j=n_K, \ \ n_j+l(\cv_{j})>l(Q),
\label{eq:parsing tail}
}
for $j \in [K:V]$ and no codeword  outside the set $\{\cv_{j}\}_{K}^V$ satisfies \eqref{eq:parsing tail}.
In other words, a parsing of a query is a covering with no overlapping over the codewords, apart from the tail of the query. 
In the tail of the query, codewords start from the same position $n_K$ and overflow the end of the sequence. 
The parsing between $K$ and $V$ contains all codewords that contain $Q_{n_K}^{l(Q)}$ as a prefix.
The string $Q_{n_K}^{l(Q)}$ is referred to as the tail of the query. 
Our interest in the PCP-SDSM problem is motivated by the next theorem.

\begin{thm}
\label{thm:PCP-SDSM}
	In the 	PCP-SDSM problem, the following holds:
	\noindent
	$-$  no retrieval error occurs,
	
	\noindent
	$-$  the minimum covering cost is always finite,
	
	\noindent
	$-$  there exists only one parsing of any query, thus this parsing is the optimal covering,
	
	\noindent
	$-$ the number of entries in the posting table is always equal to $N$.
\end{thm}

\section{Relevant Examples}
\label{sec:examples}
In the remainder of the paper, we  evaluate the efficiency for two example codes. In both cases, we consider the scenarios of binary i.i.d. sources and queries distribution. In particular, the source  distribution is obtained as 
\ea{
X^N \sim \iid \Bcal(p)^N.
\label{eq:binary sett}
}

%
\noindent
{\bf PCP-SDSM problem with a $k$-gram code:} 
Perhaps the simplest choice of posting code for the binary i.i.d. setting is the case in which $\Ccal$ contains all possible binary sequences of length $k$ such that $M=2^{k}$. 
Such a posting code is usually referred to as $k$-gram code and is typical employed in genomic research for indexing DNA sequences, such as in the well-known BLAST algorithm \cite{altschul1990basic}.

In the regime of large blocklength, the length of the posting table is obtained by constructing a Markov chain with $M$ states, each corresponding to a possible $k$-gram.
%
%
The $i$-th window of the source sequence, $K_i=X_{i}^{i+k}$,  can be represented as a state in the Markov chain: as the window slides by one position, yielding $K_{i+1}=X_{i+1}^{i+k+1}$, this corresponds to a state transition of the Markov chain.
The length of the posting sequence of each codeword in the codebook can then be obtained as the average time spent in the corresponding state of the Markov chain.
By considering the structure of the Markov chain and transition probability matrix, we obtain the steady state distribution 
$\pi_{i} =  \po^{k-w(\cv_{i})}p^{w(\cv_{i})}$ for $i=1,2\ldots,M$ and the average time spent in the state $K_i=\cv_i$.
Let us next consider the cost of each query:  let us assume that the queries are obtained as $\sum_l P_L(l) P_{Q|L=l}$ and that  
\ea{
Q|l  \sim \iid \Bcal(q)^l,
\label{eq: cond query distribution}
}
that is, given that the query length is $l$, the query is an i.i.d. sequence of Bernoulli distribution with parameter $q$ of length $l$.
The RV determining the length of the query can always be expressed as quotient and remainder of the division by $k$, i.e.
\ea{
L=k Z+R,
\label{eq:remainder and tail}
}
where $Z \in \Nbb$ and $R=[0:k-1]$. The minimum expected cost of the query is then
\ea{
\Ebb\lsb  c^*\lb \Scal(\qv) \rb \rsb=\Ebb[c(Q_1^k)] \lb Z+1_{\{R>0\}} 2^{k-R} \rb,
\label{eq:expect opt cost}
}
that is, the cost of the query is the cost of parsing the query with $Z$ $k$-grams along with covering the tail and accounting for its cost.
If the tail has length zero, than the cost of the tail is zero, otherwise the tail of the query is composed of all $2^{k-R}$  codewords with prefix $Q_{k Z}^{kZ+R}$.

\begin{lem}
\label{lem: k mer}
For the PCP-SDSM problem with a $k$-gram posting code and source and query distribution in \eqref{eq:binary sett} and \eqref{eq: cond query distribution}, the efficiency is obtained  as
\footnotesize{
\ean{
\eta & = \lb 1+ k \log \lb \po \qo + p q\rb \Ocal(\log(N)^{-1})\rb  \lb \Ebb [Z] + \Ebb\lsb 2^{k-R} \Big | R>0 \rsb \rb,
}}
for $Z$ and $R$ are defined as in \eqref{eq:remainder and tail}.
\end{lem} 

%
\noindent
{\bf PCP-SDSM problem for Run-Length Encoding (RLE):}
RLE is  very simple form of lossless data compression to encode binary data in which one symbol occurs with much higher frequency than the other.
This coding is useful, for instance, when encoding line drawings, as the black pixels are sparse. 
For such a setting, we consider the problem of identifying a specific binary pattern that can itself be described as a set of $B$ run lengths. 
For this reason, we consider a posting code of the form 
\ea{
\cv_i=\lcb \p{
1 & i=1 \\
0.\cv_{i-1}   & i \in[2:M-1]  \\
\zeros(M-1)  & i=M,
}\rnone
\label{eq:code geom}
}
where $\zeros(x)$ is the vector of all zeros of length $x$, so that $l(\cv_i)=i$ for $i\in[1:M-1]$ and $l(\cv_M)=M-1$.
In other words, the retrieval system stores the successive occurrences of zeros before a one appears, up to length $M-1$.
%
%
%
%
%
As argued for the case of $k$-grams, the length of each entry in the posting table can obtained from the average time spent in the state $K_i=\cv_i$ in the Markov chain corresponding to the windowing of the source sequence. 
Accordingly, in the regime of sufficiently large $N$, the length of each entry in the posting sequence converges to
\ean{
\lim_{N \goes \infty}|\Tcal(\cv_i,X^N)| = 
\begin{cases}
p \po^{i-1} & 1 \leq i < M \\
\po^{M-1}  & i = M,
\end{cases}
}
since each codeword apart from $\cv_M$ has unitary weight.
%
Let us next define the distribution queries: queries are of the form $S_1.S_2\ldots S_B$, where $S_j$ is a run length of length $k_j$
and  $B$ is the number of run lengths. 
The distribution of the $S_j$ is obtained 
\ea{
\Pr[ S_j=\zeros(k_j-1).1]=q \qo^{k_j-1}.
\label{eq:S_j}
}
Similar to \eqref{eq:remainder and tail},  the length of success-run $S_j$ can be expressed as quotient and remainder of the division by $M-1$, i.e. 
\ea{
l(S_j)= Z_j(M-1)+R_j,
\label{eq:remainder and tail geom }
}
so that the cost of the query is  $Z_j$-times the cost of the all zero codeword plus the cost of the codeword equal to $R_j$.
%
%
%

\begin{lem}
\label{lem: run length coding}	
For the PCP-SDSM problem with a RLE code and source and query distribution in \eqref{eq: cond query distribution} and \eqref{eq:S_j}, the efficiency is obtained  as
\footnotesize{
\ea{
\eta & =\lb 1+ \log \lb \displayfrac{pq}{(1-\po \qo)(1-\qo^{M-1})} \rb \Ocal(\log(N)^{-1}) \rb \Ebb[B].
\label{RLE efficiency}
}}
where $B$ is  the number of run lengths in the query. 
\end{lem}

\section{Numerical Evaluations}
\label{sec:Numerical Evaluations}

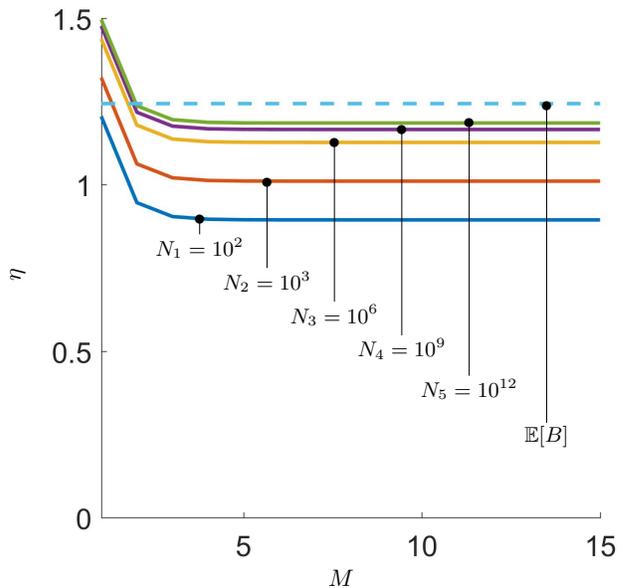
\begin{figure}
			\begin{center}
								\resizebox{0.55\textwidth}{!}{%
				\begin{tikzpicture}
				\node at (0,0.1) {\includegraphics[trim={6.2cm 0 0 0},clip,scale=0.18]{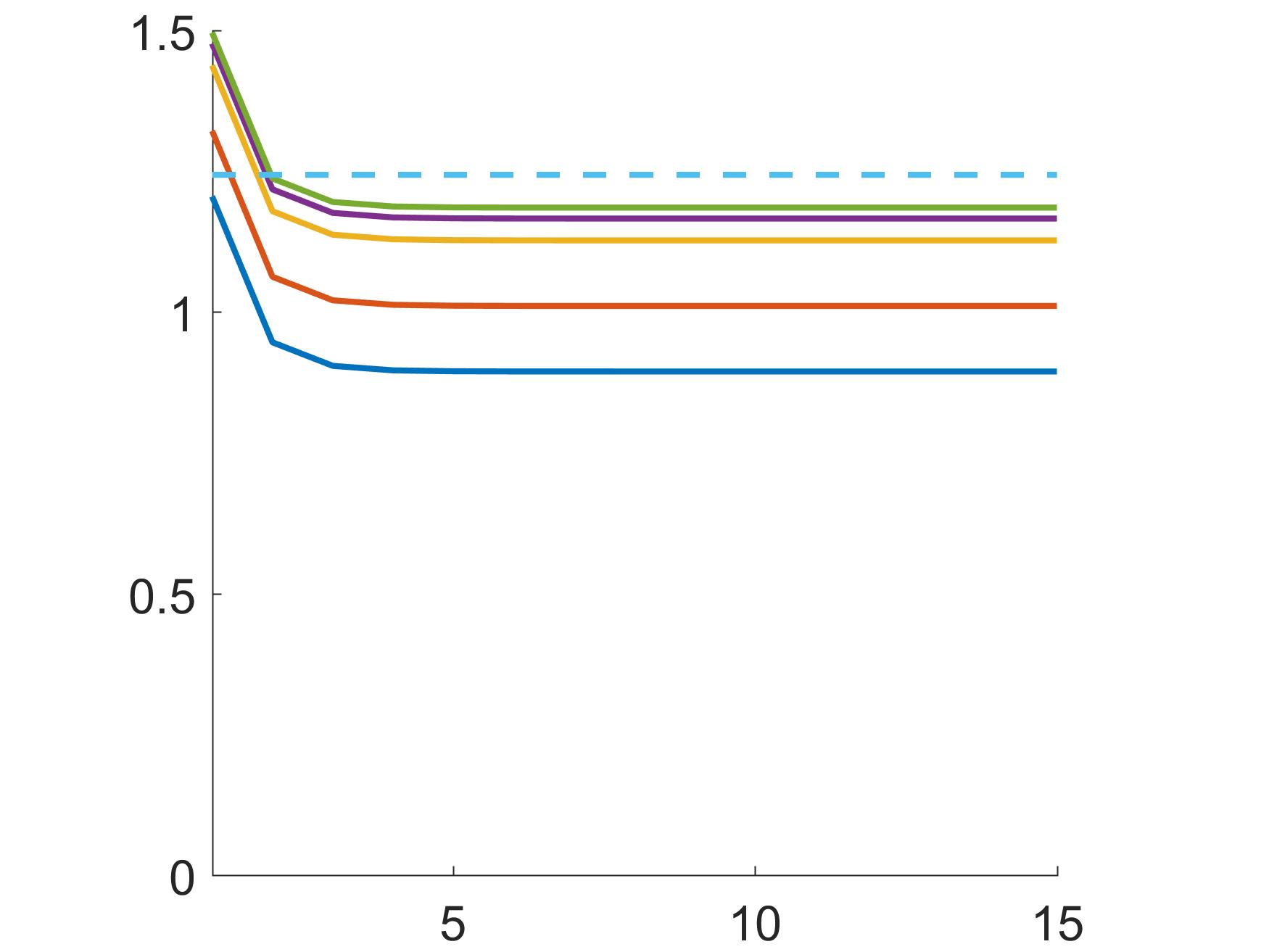}};
				\node at (-0.7,-4.3) {$M$};
				\node at (-5.5,0.2) [rotate=90]{$\eta$};
				\draw (2.35,-2.0) -- (2.35,+2.67);
				\node at (2.35,-2.2) {\small $\Ebb[B]$};
				\node at (2.35,+2.7) {$\bullet$};
				\draw (1.2,-1.3) -- (1.2,+2.47);
				\node at (1.2,-1.5) {\small $N_5 = 10^{12}$};
				\node at (1.2,+2.45) {$\bullet$};
				\draw (0.2,-0.7) -- (0.2,+2.37);
				\node at (0.2,-0.9) {\small $N_4 = 10^{9}$};
				\node at (0.2,+2.35) {$\bullet$};
				\draw (-0.8,-0.2) -- (-0.8,+2.15);
				\node at (-0.8,-0.4) {\small $N_3 = 10^{6}$};
				\node at (-0.8,+2.15) {$\bullet$};
				\draw (-1.8,+0.3) -- (-1.8,+1.57);
				\node at (-1.8,+0.1) {\small $N_2 = 10^{3}$};
				\node at (-1.8,+1.57) {$\bullet$};
				\draw (-2.8,+0.8) -- (-2.8,+1.02);
				\node at (-2.8,+0.6) {\small $N_1 = 10^{2}$};
				\node at (-2.8,+1.02) {$\bullet$};
				\end{tikzpicture}
			}
								\vspace{-0.5 cm}
				\caption{Convergence of run-length code efficiency to $\Ebb[B]$ as $N$ grows large
				}
							\vspace{-0.85 cm}
				\label{fig:run length conv}
			\end{center}
			\end{figure}

We performed two of sets of preliminary, small-scale simulations to evaluate the the proposed model and to gain some insight into the performance of the codes discussed in the previous section.
For the first simulation, prefix-free codes with maximum codeword lengths of 8 were generated by probabilistic splitting of nodes in the code trees. 
The efficiency of these randomly generated prefix-free codes was compared with that of $k$-gram codes, with a maximum $k$ of 8 and two sets of $p$ and $q$. 
We observe that values for $M$ in the neighborhood of 128, the prefix-free codes have a better efficiency than the $k$ gram for $k=7$ as the minimum expected costs are lower in this region.  
%
%
We also see that, as $M$ approaches 256, the efficiency of the prefix-free codes approaches that of the $k$-gram for $k=8$.
In the case of run-length codes, from Lem. \ref{lem: run length coding}  we see that as the source sequence grows large, the efficiency approaches the expected number of run-lengths in the query.
This behavior can be observed in Fig. \ref{fig:run length conv}. Queries with up to 4 run-lengths ($B=4$), were simulated and the distribution of $B$ was chosen to be geometric with a success probability of $0.8$.

		\begin{figure}
			\begin{center}
				\resizebox{0.55\textwidth}{!}{%
				\begin{tikzpicture}
				\node at (-0.7,0.1) {\includegraphics[trim={6.5cm 0 0 0},clip,scale=0.18]{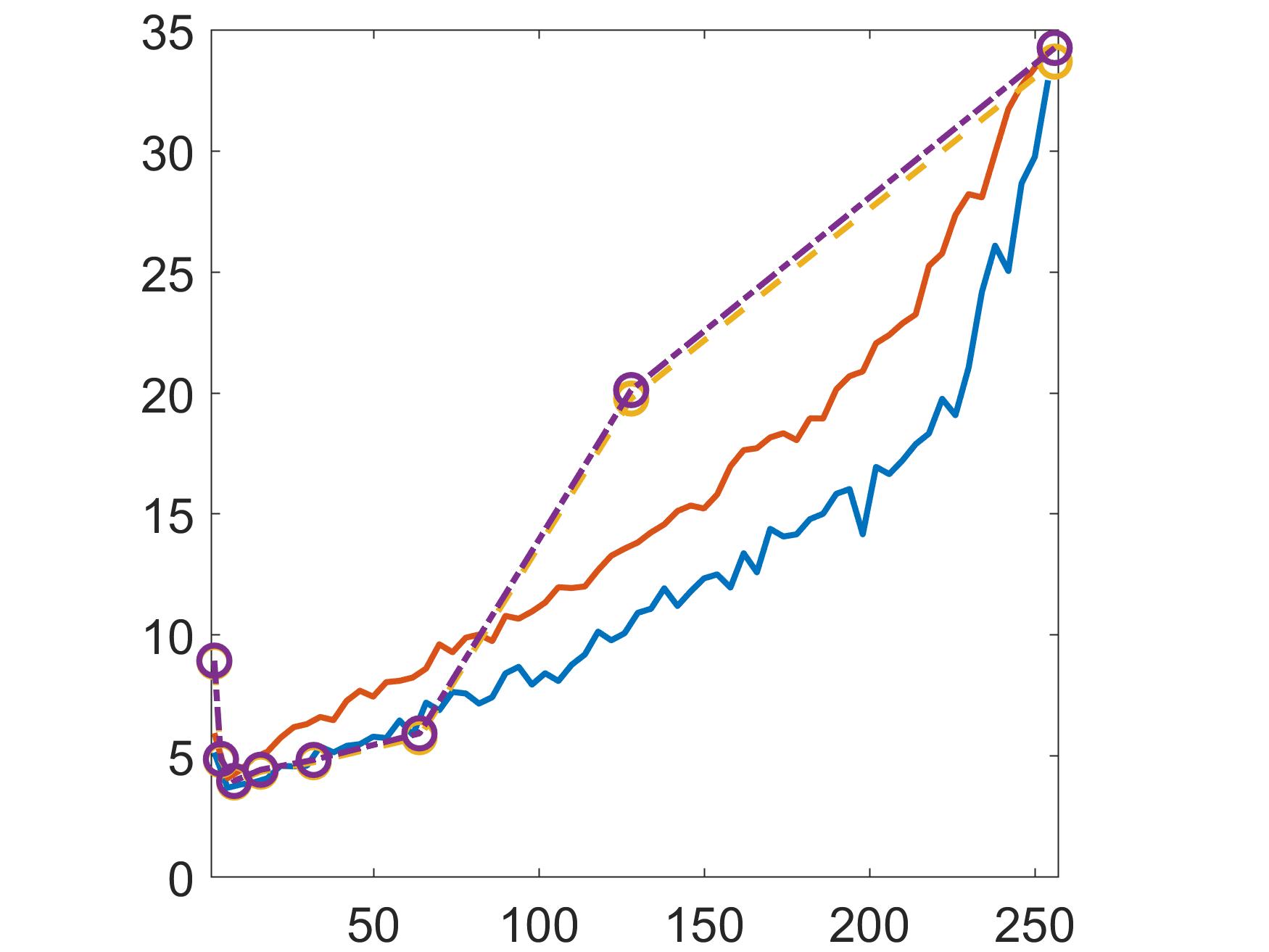}};
				\node at (-1.0,-4.3) {$M$};
				\node at (-6.2,0.2) [rotate=90]{$\eta$};
				\node at (-3.2,0.4) {\small $k=6$};
				\draw (-3.2,0.2) -- (-3.2,-2.0);
				\draw (-1.33,+0.75) -- (-1.33,2.5);
				\node at (-1.33,+2.7) {\small $k=7$};
				\draw (-4.10,-2.30) -- (-4.1,-0.7);
				\node at (-4.10,-0.5) {\small $k=5$};
				\node at (-0.7,-0.2) {$\bullet$};
				\draw (-0.7,-1.5) -- (-0.7,1.4);
				\node at (-0.7,1.4) {$\bullet$};
				\node at (-0.7,-1.7) {\small $p=0.7$};
				\node at (-0.7,-2.1) {\small $q=0.2$};
				\node at (1.05,+0.17) {$\bullet$};
				\draw (1.05,-0.5) -- (1.05,+2.60);
				\node at (1.05,-0.7) {\small $p=0.4$};
				\node at (1.05,-1.1) {\small $q=0.6$};
				\node at (1.05,+2.61) {$\bullet$};
				%
				\end{tikzpicture}			}
				\vspace{-0.5 cm}
				\caption{Comparison of efficiency of $k$-gram codes and randomly generated prefix-free codes.				
								\vspace{-0.5 cm}		}
				\label{fig:kgram and pcp code}
			\end{center}
			\end{figure}
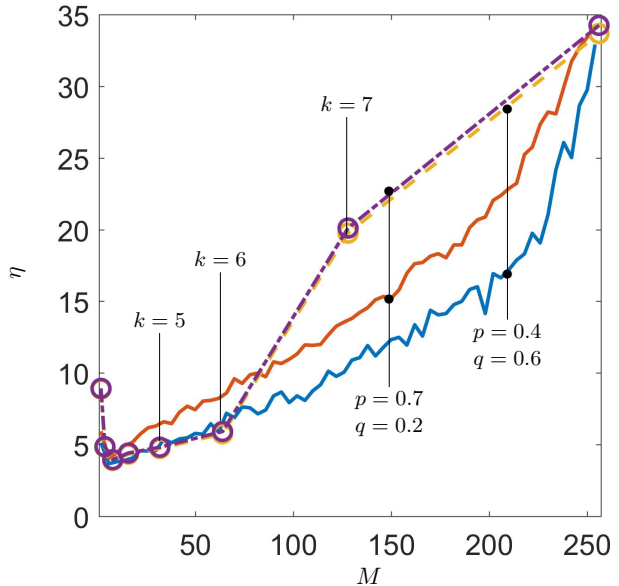

\section{Conclusion}
\label{sec:Conclusion}
In the paper we propose a statistical and information-theoretic formulation  of the dictionary-based string matching (SDSM) problem.
In the SDSM problem, a retrieval system has access to a source sequence and  it stores the position of a certain number of strings, in a table called the posting table.
Upon receiving a query from a user, the retrieval system access the entries in the table to efficiently determine the position of the matches in the source sequence.
For this problem, we assume that source and query distributions are described as random processes and we propose a cost function for the query retrieval.
Through this formulation, we are able to define an optimal posting code as the code which attains the smallest expected cost in retrieving a query.
%
For the proposed model, we provide some relevant examples and preliminary numerical evaluations.

\bibliographystyle{plain}
\bibliography{posting_bib}

\begin{thebibliography}{10}

\bibitem{altschul1990basic}
Stephen~F Altschul, Warren Gish, Webb Miller, Eugene~W Myers, and David~J
  Lipman.
\newblock Basic local alignment search tool.
\newblock {\em Journal of molecular biology}, 215(3):403--410, 1990.

\bibitem{anh2005inverted}
Vo~Ngoc Anh and Alistair Moffat.
\newblock Inverted index compression using word-aligned binary codes.
\newblock {\em Information Retrieval}, 8(1):151--166, 2005.

\bibitem{apostolico1997pattern}
Galil Apostolico.
\newblock {\em Pattern matching algorithms}.
\newblock Oxford University Press on Demand, 1997.

\bibitem{baeza2004fast}
Ricardo Baeza-Yates.
\newblock A fast set intersection algorithm for sorted sequences.
\newblock In {\em Annual Symposium on Combinatorial Pattern Matching}, pages
  400--408. Springer, 2004.

\bibitem{baeza1998fast}
Ricardo Baeza-Yates and Gonzalo Navarro.
\newblock Fast approximate string matching in a dictionary.
\newblock In {\em spire}, page 0014. IEEE, 1998.

\bibitem{baeza1999modern}
Ricardo Baeza-Yates, Berthier Ribeiro-Neto, et~al.
\newblock {\em Modern information retrieval}, volume 463.
\newblock ACM press New York, 1999.

\bibitem{baeza2003three}
Ricardo Baeza-Yates and Felipe Saint-Jean.
\newblock A three level search engine index based in query log distribution.
\newblock In {\em International Symposium on String Processing and Information
  Retrieval}, pages 56--65. Springer, 2003.

\bibitem{baeza2005experimental}
Ricardo Baeza-Yates and Alejandro Salinger.
\newblock Experimental analysis of a fast intersection algorithm for sorted
  sequences.
\newblock In {\em International Symposium on String Processing and Information
  Retrieval}, pages 13--24. Springer, 2005.

\bibitem{buttcher2016information}
Stefan B{\"u}ttcher, Charles~LA Clarke, and Gordon~V Cormack.
\newblock {\em Information retrieval: Implementing and evaluating search
  engines}.
\newblock Mit Press, 2016.

\bibitem{chen2010inverted}
David~M Chen, Sam~S Tsai, Vijay Chandrasekhar, Gabriel Takacs, Ramakrishna
  Vedantham, Radek Grzeszczuk, and Bernd Girod.
\newblock Inverted index compression for scalable image matching.
\newblock In {\em DCC}, page 525, 2010.

\bibitem{mihov2004fast}
Stoyan Mihov and Klaus~U Schulz.
\newblock Fast approximate search in large dictionaries.
\newblock {\em Computational Linguistics}, 30(4):451--477, 2004.

\bibitem{navarro2001indexing}
Gonzalo Navarro, Ricardo~A. Baeza-Yates, Erkki Sutinen, and Jorma Tarhio.
\newblock Indexing methods for approximate string matching.
\newblock {\em IEEE Data Eng. Bull.}, 24(4):19--27, 2001.


\bibitem{yan2009inverted}
Hao Yan, Shuai Ding, and Torsten Suel.
\newblock Inverted index compression and query processing with optimized
  document ordering.
\newblock In {\em Proceedings of the 18th international conference on World
  wide web}, pages 401--410. ACM, 2009.

\bibitem{yeung2012first}
Raymond~W Yeung.
\newblock {\em A first course in information theory}.
\newblock Springer Science \& Business Media, 2012.

\end{thebibliography}
\newpage
%
\subsection*{Appendix A: Proof of Th.  \ref{thm:PCP-SDSM}}
%
Any (posting) code can be represented by a $|\Xcal|$-ary tree in which  codewords are represented as nodes in the tree and each branch outgoing from an edge is labeled with one of the elements in $\Xcal$. 
The codeword associated with each node is obtained as the sequence of labeled visited in the path from the root of the three to the node.
%
%
A  prefix-free code is represented by a tree in which all codewords are leaves. A code is complete if all the leaves in the three representing the code are codewords.
Let us prove each of the properties of the PCP SDSM in Th. \ref{thm:PCP-SDSM}.

\noindent
$-$ {\bf no retrieval error occurs:} since the posting code is complete, any sequence can be parsed using such code. This follows because, starting from the beginning of the source sequence, the first codeword ends in a leaf of the tree. The next symbol in the source sequence will, consequently, start from the root of the coding tree and the second codeword will again end in a leaf. By repeating this argument, the desired property is shown.

\noindent
$-$  {\bf the minimum covering cost is always finite:} since queries are parsed with posting codewords and given that the posting code is complete, it follows that a parsing of a query always exists

\noindent
$-$ {\bf there exists only one parsing of any query, thus this parsing is the optimal covering:} again following from the completeness of the posting code, it follows that there exists a unique parsing of any codeword.

\noindent
$-$ {\bf the number of entries in the posting table is always equal to $N$:} at each position in the source sequence, a codeword exists. Following from the completeness of the storing function, such codeword is stored in the posting table.

\subsection*{Appendix B: Proof of Lem. \ref{lem: k mer}}
To construct the postings table, the source sequence $X^N$ is parsed into overlapping $k$-grams such that each $k$-gram has an overlap of $k-1$ bits with its adjacent $k$-grams and the positions of each $k$-gram in the sequence are recorded in postings lists. 

Let us first derive the average length of each posting list:  this can be determined by observing that the transition from a $k$-gram to its adjacent overlapping $k$-gram can be described through a Markov chain.
Consider the Markov chain with $M$ states, each corresponding to a possible $k$-gram, and each labelled with the decimal representation of the corresponding $k$-gram.
The transition between two states corresponds to the sliding of the $k$-gram of a position forward, i.e.  $K_i=X_{i}^{i+k}$ to $K_{i+1}=X_{i+1}^{i+k+1}$. 
The transition matrix can be constructed by observing that 
\[
P_{ij}= 
\begin{cases}
\po, & \text{if } i \leq 2^{k-1} \text{ and } j = 2i-1\\
p,   & \text{if } i \leq 2^{k-1} \text{ and } j = 2i\\
\po, & \text{if } i > 2^{k-1} \text{ and } j = 2(i-2^{k-1})-1\\
p,   & \text{if } i > 2^{k-1} \text{ and } j = 2(i-2^{k-1})\\
0,   & \text{otherwise },	
\end{cases}
\]
since the sliding removes the most significant bit in the $k$-grams and introduces a least significant bit. 
The transition probability matrix can be represented has a block matrix structure has 
\begin{equation}
P =  \begin{pmatrix}
\po & p & 0 & 0 & \cdots & 0 & 0 & 0 & 0 \\
0 & 0 & \po & p & \cdots &0 & 0 & 0 & 0 \\
\vdots  & \vdots & \vdots & \ddots & & \vdots  & \vdots & \vdots & \vdots \\
0 & 0 & 0 & 0 & \cdots & \po & p & 0 & 0 \\
0 & 0 & 0 & 0 & \cdots & 0 & 0 & \po & p \\
\po & p & 0 & 0 & \cdots & 0 & 0 & 0 & 0 \\
0 & 0 & \po & p & \cdots & 0 & 0 & 0 & 0 \\
\vdots  & \vdots & \vdots & \ddots & & \vdots  & \vdots & \vdots & \vdots \\
0 & 0 & 0 & 0 & \cdots & \po & p & 0 & 0 \\ 
0 & 0 & 0 & 0 & \cdots & 0 & 0 & \po & p \\
\end{pmatrix}.
\end{equation}
In order to calculate the average length of the posting list of a given $k$-gram, we make use of some results on Markov chains. 
For a Markov chain with transition matrix $P$ and initial state $i$, the steady state distribution is denoted by $\pi$, so that the following holds
\ea{
	\pi^{T}P = \pi^{T}.
	\label{eq:steady distribution}
}
The number of visits to state $i$ before time $n$ is 
\ea{ V_{i}(n) = \sum_{t = 0}^{n-1} 1_{\lcb X_{t}=i \rcb},
}
Under suitable conditions, we can determine the average time spent in a state using the following result  
\ea{
\Pr \lb \limni \displayfrac{V_{i}(n)}{n} =  \pi_{i} \rb = 1,
} 
and so the average length of the postings list of codeword $i$ is given by
\ea{
\Ebb[|\Tcal(\cv_i,X^N)|] = N\pi_{i}.
}
The steady-state distribution is obtained as the left eigenvector corresponding the eigenvalue at one.
We denote component $i$ of the eigenvector $u$ by $u_{i}$. This eigenvector has an eigenvalue of 1. Based on the structure of the transition matrix $P$, we guess the following
\ea{
	u_{i} =   \lb \displayfrac{\po}{p} \rb ^{k-w_{i}} \quad  i = 1,2..,M
	\label{eq:guess}
}
where $|w_{i}|$ is the hamming weight of the k-gram $w_{i}$.
If $u$ is the eigenvector corresponding to eigenvalue 1, it will satisfy the following equation
\ea{
	u^{T}P = u^{T}.
	\label{eq:eigenvector}
}
If \eqref{eq:eigenvector} holds for the choice in \eqref{eq:guess}, this must indeed be the eigenvector as eigenvectors are unique. 
Eq. \ref{eq:eigenvector} can be expressed component wise as
\ea{
	u_{j} = \sum_{i=1}^{M} u_{i}P_{ij} \quad j=1,2,..,M
}
Observe that there are only two non-zero entries in every column of $P$ and that these entries are always separated by $2^{k-1}$ rows. This implies that $w_{i+2^{k-1}} = w_{i}+1$. Using this observation, the RHS of the preceding equation for $j=0,2,..,M$ can be written as
\ea{
	=\po \lb \displayfrac{\po}{p} \rb ^{k-w_{\frac{j+1}{2}}}  +  \po\lb \displayfrac{\po}{p} \rb ^{k-w_{\frac{j+1}{2}+2^{k-1}}}  \\
	=\po \lb \displayfrac{\po}{p} \rb ^{k-w_{\frac{j+1}{2}}}  +  \po\lb \displayfrac{\po}{p} \rb ^{k-w_{\frac{j+1}{2}}-1} 
}
\ea{
	=\lb \po+\po \f p {\po} \rb  \lb \displayfrac{\po}{p} \rb ^{k-w_{\frac{j+1}{2}}}  
}
\ea{
	= \lb \displayfrac{\po}{p} \rb ^{k-w_{\frac{j+1}{2}}} = u_{\frac{j+1}{2}}=u_j
}
%
%
When $j=1,3,..,M-1$, the multiplication factor $(1-p)$ is replaced with $p$.
\ea{
 (p) \lb \displayfrac{\po}{p} \rb ^{k-w_{\frac{j-1}{2}}} + (p) \lb \displayfrac{\po}{p} \rb ^{k-w_{\frac{j-1}{2}}-1}
}
\ea{
= (p) \lb 1 +\displayfrac{p}{\po} \rb \lb \displayfrac{\po}{p} \rb  ^{k-w_{\frac{j-1}{2}}}
}
\ea{
=  \lb \displayfrac{\po}{p} \rb ^{k-w_{\frac{j-1}{2}}-1} = u_{\frac{j-1}{2}+2^{k-1}} = u_{j}
}
For the last step, note that $w_{\frac{j-1}{2}} = w_{j} - 1$ and so $w_{\frac{j-1}{2} + 2^{k-1}} = w_{j}$.

The eigenvector needs to be normalized to make it a steady-state distribution. To do this end, we calculate the sum $S$ of the elements of the eigenvector as follows
\ea{
S = \sum_{m=0}^{k} \binom{k}{k-m} (-1)^{k-m} \lb \displayfrac{\po}{p} \rb ^{k-m}
}
\ea{
 = \sum_{m=0}^{k} \binom{k}{k-m} \lb \displayfrac{\po}{p} \rb ^{k-m} = \displayfrac{1}{p^{k}}
}
Dividing $u$ by $S$ we obtain the steady-state distribution $\pi_{i}  =  (\po)^{k-w_{i}}p^{w_{i}} $
which yields
\ea{
\lim_{N \goes \infty}|\Tcal(\cv_i,X^N)| 
 \stackrel{\Pcal}{=} N(\po)^{k-w_{i}}p^{w_{i}},  
\label{eq: expect cost}
}
for $i=1,2, \ldots,M$ and where $\stackrel{\Pcal}{=}$ indicates equality in probability.
Note that the average time spent in a state only depends on the Hamming weight of the state.
%
%
%
Next we move to the analysis of servicing a query. A query $\qv$ with length $l(\qv)$ is parsed into successive $k$-grams to service it using the inverted index of $k$-grams. 
Since, in a PCP-SDSM problem there exists only one parsing, this parsing is also optimal. 
The number of $k$-grams in the parsing of length $L$ in \eqref{eq:remainder and tail} is $Z+1_{\{R>0\}} 2^{k-R}$ since the number of codewords to cover a tail of length $R$ is $2^{k-R}$.
For each $k$-grams, since symbols in a $k$-grams are iid, we conclude that \eqref{eq:expect opt cost} holds. 
Finally, we have that $Q_1^K$ is a Binomial random varible so that 
\ea{
\Ebb\lsb c\lb Q_{1}^k \rb \rsb=N \po^k M_W\lb q, k,\f {p} {\po}\rb.
}
where we have used \eqref{eq: expect cost} and where $M_W(q,k,t)$ indicates the moment generating function of a Binomial random variable with parameters $q$ and $k$ with independent variable $t$.
%
%
%
%

\subsection*{Appendix C: Proof of Lem. \ref{lem: run length coding}	}
Also, for the case of RLE, the structure of the codes allows obtaining a compact expression for the average cost of servicing a query. 
As argued for the case of $k$-grams, the length of each entry in the posting table can obtained from the average time spent in the state $K_i=\cv_i$ in the Markov chain corresponding to the windowing of the source sequence. 
\ean{
\lim_{N \goes \infty} \f 1 N \Ebb[|\Tcal(\cv_i,X^N)|] \stackrel{\Pcal}{=}  
\begin{cases}
p \po^{i-1} & 1 \leq i < M \\
\po^{M-1}  & i = M.
\end{cases}
}
%
The expected cost of each run length is then obtained as 
\ea{
\Ebb[c(S_j)] = N,
}

Since the source symbols are iid, the expected cost of a run length, decomposed as in  is 
\ea{
c(S_j)= N^{Z_j+1} p \po^{Z_j(M-1)+R_j-1} ,
}

In this case, we normalize the cost term to factor out the effect of the sequence length, and then take the expected value 
\ea{
\Ebb [c(S_j)]=\sum_{Z_j = 0}^{\infty} \sum_{R = 1}^{M-1} pq (\po \qo) ^{Z_j(M-1)+R_j-1}\\
=\displayfrac{pq}{\po \qo} \sum_{Z_j = 0}^{\infty} (\po \qo) ^{Z_{j}(M-1)} \sum_{R = 1}^{M-1} (\po \qo) ^{R}\\
= \displayfrac{pq}{(1 - \po\qo)(1-\qo^{M-1})}.}

This is generalized to the case of $b$ run lengths by using the fact that run lengths are i.i.d.
\ea{
\Ebb[c(\qv)|B = b] = \lb \displayfrac{pq}{(1 - \po\qo)(1-\qo^{M-1})} \rb^{b}.}

Taking the log and re-normalizing, we obtain the expression \eqref{RLE efficiency}

\end{document}